\documentclass[superscriptaddress,twocolumn,prb,showpacs]{revtex4}
\usepackage[dvips]{graphicx}
\usepackage{amsmath}

\def\beq{\begin{equation}}
\def\eeq{\end{equation}}
\def\bea{\begin{eqnarray}}
\def\eea{\end{eqnarray}}
\def\nn{\nonumber\\}
\def\r{{\bf r}}

\def\v{{\bf v}}
\def\k{{\bf k}}
\def\R{{\bf R}}
\def\G{{\bf G}}

\def\K{{\bf K}}
\def\A{{\bf A}}

\def\M{{\bf M}}

\def\0{{\bf 0}}

\def\j{{\bf j}}

\def\f{\tilde{\bf f}}
\def\g{\tilde{\bf g}}
\def\h{\tilde{\bf h}}

\def\sLC{_{\rm LC}}
\def\sIC{_{\rm IC}}

\def\da{\partial_{\alpha}}
\def\db{\partial_{\beta}}

\def\park{\partial_{\bf k}}
\def\ket#1{\vert#1\rangle}
\def\bra#1{\langle#1\vert}
\def\me#1#2#3{\langle#1\vert#2\vert#3\rangle}
\def\ev#1{\langle#1\rangle}

\newcommand{\emi}[1]{{\rm e}^{-i #1}}
\newcommand{\ei}[1]{{\rm e}^{i #1}}
\newcommand{\equ}[1]{Eq.~(\ref{#1})}
\newcommand{\eqs}[2]{Eqs.~(\ref{#1}) and (\ref{#2})}
\newcommand{\intk}{\int_{\rm BZ} \!\!\!\! d{\bf k} \;}
\newcommand{\intkm}{\int_{\epsilon_{n\k} \leq \mu} \!\!\!\! d{\bf k} \;}
\def\kab{_{\k,\alpha\beta}}

\def\wda{\widetilde{\partial}_{\alpha}}
\def\wdb{\widetilde{\partial}_{\beta}}

\def\wdi{\widetilde{\partial}_i}
\def\wdj{\widetilde{\partial}_j}
\def\wdl{\widetilde{\partial}_l}
\def\wu{\widetilde{u}}
\def\q{{\bf q}}
\def\nkpbi{_{n,\k+{\bf b}_i}}
\def\nkmbi{_{n,\k-{\bf b}_i}}
\def\nkpq{_{n,\k+\q}}
\def\F{{\bf F}}

\begin{document}

\title{Orbital magnetization in crystalline solids:\\
Multi-band insulators, Chern insulators, and metals}
\author{Davide Ceresoli}
\affiliation{Scuola Internazionale Superiore di Studi Avanzati (SISSA/ISAS)
and DEMOCRITOS, via Beirut 2-4, 34014 Trieste, Italy}
\author{T. Thonhauser}
\affiliation{Department of Physics and Astronomy, Rutgers University,
Piscataway, New Jersey 08854, USA}
\author{David Vanderbilt}
\affiliation{Department of Physics and Astronomy, Rutgers University,
Piscataway, New Jersey 08854, USA}
\author{R. Resta}
\affiliation{Dipartimento di Fisica Teorica Universit\`a di Trieste and
DEMOCRITOS, strada Costiera 11, 34014 Trieste, Italy}

\begin{abstract}

We derive a multi-band formulation of the orbital magnetization in a
normal periodic insulator (i.e., one in which the Chern invariant,
or in 2d the Chern number, vanishes).  Following the approach used
recently to develop the single-band formalism [T. Thonhauser, D.
Ceresoli, D. Vanderbilt, and R. Resta, Phys. Rev. Lett. {\bf 95},
137205 (2005)], we work in the Wannier representation and find that
the magnetization is comprised of two contributions, an obvious one
associated with the internal circulation of bulk-like Wannier
functions in the interior and an unexpected one arising from net
currents carried by Wannier functions near the surface.  Unlike the
single-band case, where each of these contributions is separately
gauge-invariant, in the multi-band formulation only the \emph{sum}
of both terms is gauge-invariant. Our final expression for the
orbital magnetization can be rewritten as a bulk property in terms
of Bloch functions, making it simple to implement in modern code
packages. The reciprocal-space expression is evaluated for 2d
model systems and the results are verified by comparing to the
magnetization computed for finite samples cut from the bulk.
Finally, while our formal proof is limited to normal insulators, we
also present a heuristic extension to Chern insulators (having
nonzero Chern invariant) and to metals. The validity of this
extension is again tested by comparing to the magnetization of
finite samples cut from the bulk for 2d model systems. We find
excellent agreement, thus providing strong empirical evidence in
favor of the validity of the heuristic formula.

\end{abstract}
\maketitle

\section{Introduction}

During the last decade, charge and spin transport phenomena in magnetic
materials and nanostructures have attracted much interest due to their
important role for spintronic devices.\cite{fabian04} An adequate
description of magnetism in these materials, however, should not only
include the spin contribution, but also should account for effects
originating in the orbital magnetization. In light of this, it is
surprising that the theory of orbital magnetization has long remained
underdeveloped. Earlier attempts to develop such a theory used
linear-response methods, which allow calculations of magnetization
\emph{changes},\cite{linear, Sebastiani01, Mauri, Sebastiani02} but not
of the magnetization itself.

Just recently, a new approach using Wannier functions (WFs) has been
proposed,\cite{ChemPhysChem,Thonhauser05} which nicely parallels the
analogous case of the electric polarization. The primary difficulty in
both cases is that the position operator $\r$ is not well-defined in
the Bloch representation. Since WFs are exponentially localized in an
insulator, this difficulty disappears if the problem is reformulated in
the Wannier representation. For the polarization, this approach lead to the
development of the modern theory of polarization in the early
1990s.\cite{KSV,rap-a12} Similarly, in the case of the orbital
magnetization, where the circulation operator ${\r}\times{\v}$ is
ill-defined in the Bloch representation, the Wannier representation was
used to derive a theory for the orbital magnetization of periodic
insulators.\cite{Thonhauser05}

While the formalism developed in Ref.~\onlinecite{Thonhauser05}  lays a firm
foundation for the orbital magnetization, its application is limited to
certain systems, such as single-band models and insulators.  In this paper we
expand the applicability to a much wider class of systems by developing a
corresponding multi-band formalism, essential for most ``real'' materials. 
This extension is nontrivial and the corresponding proof of gauge invariance
is much more complex than for the single-band case.  We
proceed in two steps.  First, we carry out a derivation for the case
of an insulator with zero Chern invariant.  Second,
we give heuristic arguments
for an extension of our formalism to metals and Chern insulators, i.e. 
systems with a non-zero Chern invariant,
arriving at a formula identical to that proposed by
Xiao, Shi and Niu\cite{Xiao05} on the basis of semiclassical arguments.
Chern insulators have been introduced into the
theoretical literature by means of model Hamiltonians in 2d which break
time-reversal (TR) symmetry without breaking translational symmetry,
\cite{Haldane88} i.e., maintaining a vanishing {\it macroscopic}
magnetic field. Despite the absence of a macroscopic field, Chern insulators
share several properties with quantum-Hall systems, most notably
the quantization of the transverse conductivity in 2d.\cite{Haldane88}
To the best of our knowledge, there is no known experimental realization of a
Chern insulator (in zero field) in either 2d or 3d, and the search for such
a system remains a fascinating challenge.

Our extensions to metals and Chern insulators are heuristic and {\it not}
based on an analytical proof. The fact that our final formula is identical
to the one derived from the semiclassical wavepacket treatment\cite{Xiao05}
is reassuring, but neither of these approaches can yet be said to constitute
a ``derivation'' of the formula in the fully quantum context.
Nevertheless, we provide strong numerical evidence of their
validity, thus posing a theoretical challenge:  how to provide an
analytic proof of the heuristic formula, beyond the
range of the semiclassical approximation, for both the metallic and
Chern-insulating cases.

Before proceeding, we emphasize that the present work only addresses
the question of how to compute the orbital magnetization for a given
independent-particle Hamiltonian.  Many interesting questions remain
concerning which flavor of density-functional theory (DFT) or which
exchange-correlation (XC) functional might give the most accurate
orbital magnetization.  While exact Kohn-Sham (KS) density (or
spin-density) functional theory is guaranteed to yield the correct
charge (or spin) density,\cite{DFT} there is no reason to expect it to
yield the correct orbital currents.  The orbital magnetization, being
defined in terms of surface currents, is not guaranteed to be
correct either.  A prescription that seems more suited to the present
situation is that of Vignale and Rasolt,\cite{Vignale88} in which the
spin-labeled density and current $\{n_\sigma(\r),\;\j_\sigma(\r)\}$ are
connected to corresponding scalar and vector potentials
$\{V_\sigma(\r),\;\A_\sigma(\r)\}$.  However, it is an open question
whether an approximate Vignale-Rasolt XC functional exists that can
give improved values of magnetization in practice. While time-dependent
density functional theory (TDDFT) is more developed,\cite{TDDFT} this
theory only establishes a connection between $n(\r,t)$ and $V(\r,t)$,
and a knowledge of $n(\r,t)$ is only sufficient to determine the
longitudinal part of $\j(\r,t)$, not the transverse part upon which the
orbital magnetization depends.  An alternative approach worthy of
exploration is time-dependent current-density functional theory
(TD-CDFT),\cite{TD-CDFT} in which $\{n(\r,t),\;\j(\r,t)\}$ is connected
to $\{V(\r,t),\;\A(\r,t)\}$.  However, the present problem is
essentially a static problem, and it is therefore unclear whether
TD-CDFT would provide any practical advantage over the Vignale-Rasolt
theory.  Finally, it is worth remembering that even in standard DFT,
the mapping from interacting density to non-interacting potential is
sometimes pathological (e.g., a KS metal can represent an interacting
insulator).  In the present work, we bypass all these interesting
issues, and only consider how to compute the magnetization for a given
Kohn-Sham Hamiltonian arising from some
unspecified version of DFT in the context of broken TR
symmetry.

We have organized this paper as follows.  In
Sec.~\ref{sec:theory} we derive the multi-band theory of orbital
magnetization in crystalline solids. After some definitions and generalities,
we start by considering the orbital magnetization of a finite
sample. The resulting expression is then transformed to reciprocal space
and its gauge invariance is demonstrated. We then give a heuristic extension
of our formalism to metals and Chern insulators. In
Sec.~\ref{sec:nummerics}, numerical results for the orbital
magnetization are presented for several different systems. We conclude in
Sec.~\ref{sec:conclusions}.
Some details concerning the finite-difference evaluation
of the magnetization and certain properties of the nonAbelian Berry
curvature are deferred to two appendices.

\section{Theory}
\label{sec:theory}

\subsection{Generalities}
Our basic starting point
is a single-particle KS Hamiltonian~\cite{DFT} having the translational
symmetry of the crystal, but having no TR symmetry: as said above,
translational symmetry of the Hamiltonian implies vanishing of the {\it
macroscopic} magnetic field. There may, however, be a microscopic magnetic
field ${\bf B}$ that averages to zero over the unit cell, and we assume that
a particular magnetic gauge has been chosen once and for all to represent
this magnetic field.  Wavevector $\k$ is a good quantum number under these
conditions.  This could be realized, for example, in systems in which the TR
breaking comes about through the spontaneous development of ferromagnetic
order or via spin-orbit coupling to a background of ordered local
moments.\cite{Haldane88, Ohgushi00, Jungwirth02, Murakami03, Yao04} Notice
that we carefully avoid referring to an {\it externally applied} field; such
concept is legitimate only for a finite sample, free-standing
in vacuo. Indeed, for a finite sample, the relationship between the externally
applied field and the ``internal'' (or screened) one depends on the sample
shape. For an extended sample in the thermodynamic limit, the only
legitimate and measurable field is the screened ${\bf B}$ field which is
present inside the material. In the present work, the cell-average of this
field is assumed to vanish.  


As usual, we let $\epsilon_{n\k}$ and $\ket{\psi_{n\k}}$ be the Bloch
eigenvalues and eigenvectors of $H$, respectively, and
$u_{n\k}(\r)=\emi{\k\cdot\r}\psi_{n\k}(\r)$ be the corresponding
eigenfunctions of the effective Hamiltonian $H_\k = \emi{\k\cdot\r} H
\ei{\k\cdot\r}$.  We choose to normalize them to one over the crystal cell of
volume $\Omega$.

The notation is intended to be flexible as regards the spin
character of the electrons.  If we deal with spinless electrons, then
$n$ is a simple index labeling the occupied Bloch states;
factors of two may trivially be inserted if one has in mind
degenerate, independent spin channels.  In the context of the local
spin-density approximation (LSDA), in which spin-up and spin-down electronic
states are separate eigenstates of spin-up and spin-down Hamiltonians,
one may let $n$ range over both sets of bands, but with the understanding
that inner products or matrix elements between spin-up and spin-down
bands always vanish.  Of more realistic interest here is the case of
a fully non-collinear treatment of the magnetism, as for the case of
a Hamiltonian containing the spin-orbit operator.  In this case,
$n$ labels bands that are neither purely spin-up nor spin-down,
$\ket{u_{n\k}}$ must be understood to be a spinor wavefunction, and
the contraction over spin degrees of freedom is understood to be
included in the definition of inner products like
$\ev{u_{n\k}\vert u_{n'\k}}$ and matrix elements like
$\ev{u_{n\k}\vert H_\k \vert u_{n'\k}}$.

A key issue in the present work is the additional ``gauge freedom''
in which the occupied Bloch orbitals at fixed $\bf k$ are allowed
to be transformed among themselves by an arbitrary unitary
transformation.  In fact, any KS ground-state electronic property
should be uniquely determined by the {\it subspace} of occupied
orbitals as represented by the one-particle density matrix;
the occupied orbitals just provide a convenient orthonormal
representation for this subspace.  Moreover, when it comes to the
formulation of Wannier functions (WFs) for composite energy bands,
the $n$-th WF is generally not simply the Fourier transform
of the $n$-th band of Hamiltonian eigenvectors, but instead,
of a manifold of states $\ket{u_{n\k}}$ which are related to the
eigenstates by a $\k$-dependent unitary transformation.\cite{Marzari97}
Thus, in what follows, we allow $\ket{u_{n\k}}$ to refer to this
generalized interpretation of the $n\k$ labels unless otherwise
specified.  In addition, we introduce a generalized ``energy matrix''
\beq
E_{nn'\k}=\ev{u_{n\k} \vert H_\k \vert u_{n'\k}} .
\label{energy}
\eeq
which reduces to $E_{nn'\k}=\epsilon_{n\k}\delta_{nn'}$ in the special case
of the ``Hamiltonian gauge'' in which the $\ket{u_{n\k}}$ {\it are}
eigenstates of $H_\k$.

A key quantity characterizing a three-dimensional KS insulator in absence of
TR symmetry is the (vector) Chern invariant\cite{Thouless}
\beq {\bf C} = \frac{i}{2\pi} \intk \sum_n \ev{\park u_{n\k} | \times |
\park u_{n\k}} , \label{chern} \eeq with the usual meaning of the cross
product between three-component bra and ket states. Here and in the
following the sum is over the occupied $n$'s only, the integral is over
the Brillouin zone (BZ), and $\park = \partial/\partial \k$. The Chern
invariant is gauge-invariant in the above generalized sense (as will
be shown in Sec.~\ref{sec:gauge})
and---for a three-dimensional crystalline
system---is quantized in units of reciprocal-lattice vectors $\G$.
In Secs.~\ref{sec:magfinite}-\ref{sec:gauge} we assume that we are
working with insulators with {\it zero} Chern invariant; the more
general case will be discussed only later in
Secs.~\ref{sec:metalchern}-\ref{sec:twod}.

Owing to the zero-Chern-invariant condition, the Bloch orbitals can be chosen
so as to obey $\ket{\psi_{n\k\!+\!\G}} = \ket{\psi_{n\k}}$ (the so-called
periodic gauge), which in turn warrants the existence of Wannier functions
(WFs) enjoying the usual properties. (For a Chern insulator,
it is not clear whether a Wannier representation exists.)
We shall denote as $\ket{n\R}$ the $n$'th WF in cell $\R$. These WFs
are related via
\bea \ket{u_{n\k}} &=& \sum_\R
\ei{\k \cdot(\R - \r)} \ket{n\R} , \nn \ket{n\R} &=& \frac{\Omega}{(2\pi)^3}
\intk \ei{\k \cdot (\r - \R)} \ket{u_{n\k}} , \label{wfs} \eea
to the Bloch-like orbitals $\ket{u_{n\k}}$ defined in the generalized sense
discussed just above \equ{energy}.

\subsection{The magnetization of a finite sample}
\label{sec:magfinite}

We start by considering a macroscopic sample of $N_c$ cells (with $N_c$
very large but finite) cut from a bulk insulator, having $N_b$ occupied
bands, with ``open'' boundary conditions. The finite system then has
$N \simeq N_c N_b$ occupied KS orbitals.
Suppose we perform a unitary
transformation upon them, by adopting some localization criterion. By
invariance of the trace the orbital magnetization of the finite system
is written in terms of the localized orbitals $\ket{w_i}$ as
\beq \M = -\frac{1}{2 c \Omega N_c} \sum_{i=1}^{N} \me{w_i}{\r
\times \v}{w_i} , \label{total} \eeq
where the velocity is defined as
\beq \v = i[H,\r] . \label{group} \eeq
In the case of density-functional implementations, it should be noted
that $\v$ may differ from ${\bf p}/m$ because of the presence of
microscopic magnetic fields (which introduce ${\bf p}\cdot{\bf A}$
terms in the Hamiltonian), spin-orbit interactions, or semilocal or
nonlocal pseudopotentials.  In the case of tight-binding implementations,
the matrix representations of $H$ and $\r$ are assumed to be known
($\r$ is normally taken to be diagonal) in the tight-binding basis,
and $\v$ is then defined through \equ{group}.

We divide the sample into
an ``interior'' and a ``surface'' region, in such a way that the latter
occupies a non-extensive fraction of the total sample volume in the
thermodynamic limit. The orbitals $\ket{w_i}$ which are localized in the
interior region converge exponentially to the WFs $\ket{n\R}$ of the
periodic infinite system; for instance, if the Boys\cite{Boys60}
localization criterion is adopted, they become by construction the
Marzari-Vanderbilt\cite{Marzari97} maximally localized WFs.  Therefore
the interior is composed of an integer number $N_i$ of replicas of a unit cell
containing $N_b$ WFs each. Note that this choice is not unique; there is
freedom both to shift all of the $\R$'s by some constant vector
(effectively changing the origin of the unit cell), or to shift any one of
the WFs by a lattice vector, or to carry out a unitary remixing of the
bands.  We insist only that some consistent choice is made once and for
all.

The remaining $N_s$ localized orbitals residing in the surface region
need not resemble bulk WFs; we denote them as $\ket{w_s}$ and continue
to refer to them as ``WFs" in a generalized sense.  We thus
partition the entire set of $N$ WFs of the finite sample into $N_iN_b$
ones belonging to the interior and $N_s$ ones in the surface region.
Correspondingly, the contribution to the orbital magnetization $\M$
coming from the interior orbitals will be denoted as $\M\sLC$ (for
``local circulation''), while that arising from the surface orbitals
will be referred to as $\M\sIC$ (for ``itinerant circulation'').  We
will take the thermodynamic limit in such a way that $N_s$ grows more
slowly with sample size than does $N_i$, so that
$N_s/N_i\rightarrow0$.  Because of the ambiguities discussed in the
previous paragraph, we do not expect $\M\sLC$ and $\M\sIC$ to be
separately gauge-invariant.  However, their sum, \equ{total},
must be gauge-invariant.

Since the interior orbitals are bulk-like, we have, following
\equ{total},
\beq \M\sLC  = -\frac{1}{2 c \Omega N_c} \sum_{n\R} \me{n\R}{(\r - \R)
\times \v}{n\R}, \eeq where the number of $\R$ vectors in the sum is smaller
than $N_c$ only by a nonextensive fraction, and we have used that $\sum_n
\me{n\R}{\v}{n\R} = 0$. Because of the zero-Chern-invariant condition
the WFs enjoy the usual translational symmetry, and we finally find that
\beq \M\sLC = -\frac{1}{2 c \Omega} \sum_{n}
\me{n\0}{\r \times \v}{n\0} \label{mlc} \eeq
in the thermodynamic limit.

We now consider the contribution from the $N_s$ surface orbitals, whose
centers we denote as $\r_s = \me{w_s}{\r}{w_s}$ : \beq \M\sIC = -\frac{1}{2 c
\Omega N_c} \sum_{s=1}^{N_s} ( \me{w_s}{(\r - \r_s) \times \v}{w_s} + \r_s
\times \me{w_s}{\v}{w_s} ). \eeq The first term in parenthesis clearly
vanishes in the thermodynamic limit, while the second term---owing to the
presence of the ``absolute'' coordinate $\r_s$---does not.
At first sight, this second term in $\M\sIC$ appears to depend on
surface
details; instead, we are going to prove that even this term can be recast in
terms of {\it bulk} Wannier functions. Remarkably, both $\M\sLC$ and $\M\sIC$
are genuine bulk properties in the thermodynamic limit, and can eventually be
evaluated as BZ integrals.

\begin{figure}\begin{center}
  \includegraphics[width=0.8\columnwidth]{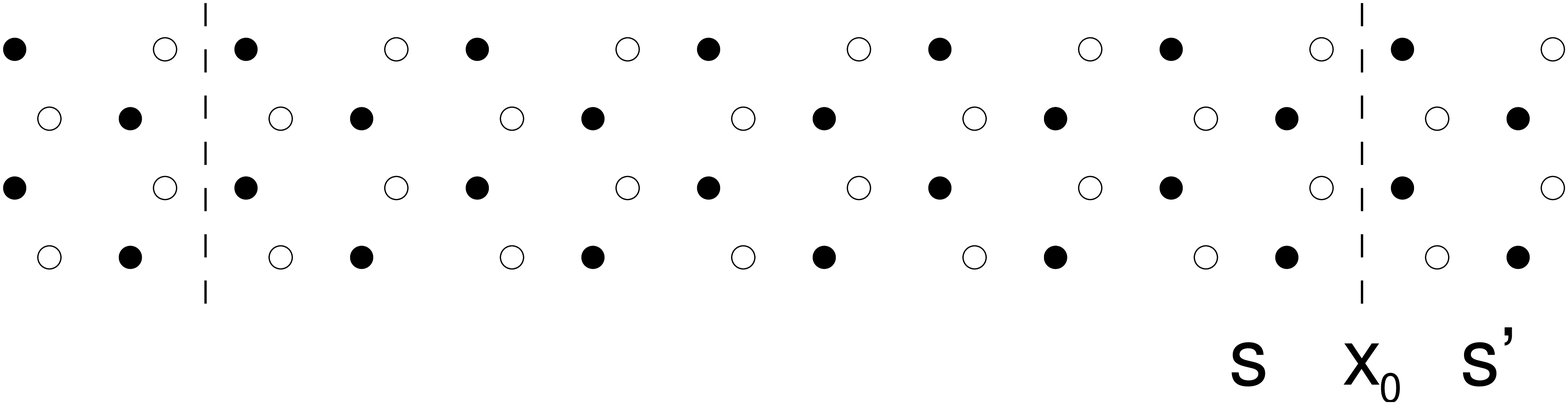}
  \caption{Horizontal slice from a sample that extends indefinitely in the
  vertical direction. Vertical dashed lines delimit bulk and surface regions
  in which WFs are labeled by $s$ and $s'$, respectively.}
  \label{fig:strip}
\end{center}\end{figure}

We consider a surface facing in the $+\hat{x}$ direction, and identify
a surface region given by $x > x_0$ as in Fig.~\ref{fig:strip}.  
There is then a contribution to the macroscopic surface current $\K$ flowing 
at the surface that is given by
\beq \K = -\frac{1}{A}
{\sum_{s'}}' \me{w_{s'}}{\v}{w_{s'}} , \label{Kdef} \eeq
where the primed sum is taken over the surface WFs whose $yz$ coordinates
are within one surface unit cell of area  $A$.  Because
$\me{w_{s'}}{\v}{w_{s'}}$ decays exponentially to zero with distance from
the surface, it is straightforward to capture the entire surface current
by letting the width of the surface region diverge slowly (say, as
the 1/4 power of linear dimension) in the thermodynamic limit, so
that $x_0$ is moved arbitrarily deep into the bulk.

It is now expedient to use the identity \beq \me{w_i}{\v}{w_i} = \sum_{j}
\v_{\ev{j,i}} , \label{donat2} \eeq where \beq \v_{\ev{j,i}} =  2 \, {\rm
Im}\, \me{w_i}{\r}{w_j} \me{w_j}{H}{w_i} \label{donat} \eeq  has the
interpretation of a current ``donated from WF $\ket{w_j}$ to WF
$\ket{w_i}$'', and exploit the fact that the total current carried by any
subset of WFs can be computed as the sum of all $\v_{\ev{j,i}}$ for which
$i$ is inside and $j$ is outside the subset.  Applying this to the piece
of surface region considered above, we get \beq \K = -\frac{1}{A}
{\sum_{s'}}' \sum_{s \neq s'} \v_{\ev{s,s'}} .  \label{piece2} \eeq
Setting the boundary deep enough below the surface to be in a bulk-like
region and invoking the exponential localization of the WFs and of related
matrix elements, we can identify $\ket{w_s}$ and $\ket{w_{s'}}$ with the
bulk WFs $\ket{m\R}$ and $\ket{n\R'}$, respectively. Exploiting
translational symmetry, $\v_{\ev{m\R,n\R'}} = \v_{\ev{m0,n(\R'-\R)}}$,
Eq.~(\ref{piece2}) becomes \beq \K = -\frac{1}{A} \,{\sum_{R_x < x_0}}\;
{\sum_{R'_x > x_0}}^{\!\!\!\prime} \sum_{mn}\v_{\ev{m0,n(\R'-\R})} \;,
\label{piece3} \eeq where the lattice sum is still restricted to the $\R'$
vectors whose $yz$ coordinates are within the surface unit cell.  The
number of terms in the lattice sum of Eq.~(\ref{piece3}) having a given
value of $\R'-\R$ is just $(R'_x - R_x) A/\Omega$ if $(R'_x-R_x)>0$ and
zero otherwise. With a change of summation index, Eq.~(\ref{piece3})
becomes
\beq \K = -\frac{1}{2\Omega} \sum_{\R} R_x \sum_{mn}
\v_{\ev{m0,n\R}} \;, \label{piece4} \eeq
where the factor of 2 enters
because the sum has been extended to all $\R$. Notice that the
surface-cell size has eventually disappeared.

Evidently the corresponding surface current on a surface with unit normal
$\hat{n}$ is then  \beq K_\alpha(\hat{n})=\sum_\beta G_{\alpha\beta} n_\beta
\eeq where \beq G_{\alpha\beta} = -\frac{1}{2\Omega} \sum_{\R} \sum_{mn}
\v_{\ev{m0,n\R},\alpha} R_\beta . \eeq  Now for a sample of size $L_x\times
L_y\times L_z$, the left and right faces carry currents of $\pm L_y L_z
G_{yx}$ separated by a distance $L_x$, and thus contribute to the magnetic
moment per unit volume as $G_{yx}/2c$; similarly, the front and back faces
contribute as $-G_{xy}/2c$. Together they contribute to $M_z$ as $-G^{\rm
A}_{xy}/c$ where  \beq G^{\rm A}_{\alpha\beta} = \frac{1}{2}\,\left(\,
G_{\alpha\beta}-G_{\beta\alpha}\,\right) , \eeq is the antisymmetric part of
the $G$ tensor.  Deriving corresponding expressions for $M_x$ and $M_y$ by
permutation of indices, the contribution of the surface current in
\equ{piece4} to the magnetization can thus be cast in a
coordinate-independent form and evaluated for the whole sample surface in the
thermodynamic limit as 
\beq \M\sIC = -\frac{1}{4c\Omega} \sum_{mn\R} \R
\times \v_{\ev{m0,n\R}} \;.  \label{mic} \eeq  Note that \equ{mic} describes the current circulating in the {\it
surface} WFs, while the expression on its right-hand side involves only {\it
bulk} WFs.  

This is quite remarkable, and indeed it is one of the central results of this
paper, as well as of Ref.~\onlinecite{Thonhauser05}. It implies that even
$\M\sIC$ is a bulk property, as anticipated above. This may appear
counterintuitive, but indeed closely parallels a well-known (and equally
counterintuitive) feature of the quantum-Hall effect, where the Hall current
is accomodated by chiral edge states.\cite{Halperin82,Yosh} Nevertheless,
these edge currents are
completely determined by bulk properties of the system, and can be
evaluated by adopting toroidal boundary conditions in which the sample has no
edges. Such a finding, in fact, is one of the most remarkable results of the
quantum-Hall theoretical
literature.~\cite{Thouless,Thouless82,Kohmoto85,Niu85} We also notice that
the bulk nature of $\M\sIC$ guarantees that our general expressions, valid in
the thermodynamic limit, apply regardless of whether surface states are
present in bounded samples, and if they are present, regardless of their
character.

It might be thought that the surface currents $\K$ must flow
parallel to the surface, and thus that the diagonal elements
$G_{xx}$ and $G_{yy}$ must vanish, or more generally, that the
symmetric component
\beq G^{\rm S}_{\alpha\beta} = \frac{1}{2}\,\left(\,
G_{\alpha\beta}+G_{\beta\alpha}\,\right) 
\label{GS}
\eeq
of the $G$ tensor should vanish.  This turns out {\it not} to be
true.  In some of our tight-binding model calculations, we have
explicitly computed the right-hand side of \equ{Kdef} and confirmed
the existence of a surface-normal component of $\K$.

The explanation is
that $\K$, as defined by \equ{Kdef}, is only one contribution
to the physical macroscopic surface current.  There is an additional
contribution arising from the fact that, when TR symmetry is broken,
the second-moment spreads\cite{Marzari97} of the WFs are {\it not}
generally stationary with respect to time.
For example, if the WFs are in the process of expanding, then
electron charge is in the process of spilling out of the surface.
To formalize this notion, we introduce a symmetric Cartesian tensor
\beq
W_{\alpha\beta}=-\frac{1}{2\Omega N_c}\sum_i\me{w_i}{r_\alpha v_\beta
    + v_\alpha r_\beta}{w_i}
\eeq
that is a kind of symmetric analog of the antisymmetric expression
for $\bf M$ given in \equ{total}.  If $W_{\alpha\beta}$ is non-zero,
then we would expect surface currents of the form
$K_\alpha(\hat{n})=\sum_\beta W_{\alpha\beta}n_\beta$.  If present,
these would violate continuity.  However, they are not present,
because we can write
\beq
W_{\alpha\beta}=-\frac{1}{2\Omega N_c}\,\frac{d}{dt}\sum_i
\me{w_i}{r_\alpha r_\beta}{w_i} \;.
\eeq
Noting that the trace of any operator (here $r_\alpha r_\beta$)
must be independent of time in any stationary state (here
the ground state of the finite sample), it follows that
$W_{\alpha\beta}=0$.  Nevertheless, if we were to follow a route
parallel to that used for the treatment of $\bf M$ earlier in
this section, we could decompose $W$ into a ``local spread''
part $W_{\rm LS}$ and an ``itinerant spread'' part $W_{\rm IS}$.
The former is
\bea
W_{{\rm LS},\alpha\beta}
&=&-\frac{1}{2\Omega}\sum_n\me{n\0}{r_\alpha v_\beta +v_\alpha r_\beta}{n\0}
\nn
&=&-\frac{1}{2\Omega}\,\frac{d}{dt}\sum_n\me{n\0}{r_\alpha r_\beta}{n\0}
\;,
\eea
which is just related to the rate of spread of the bulk WFs in
one bulk unit cell, while the latter is just
$W_{{\rm IS},\alpha\beta}=G^{\rm S}_{\alpha\beta}$ of \equ{GS}.
Because the total $W_{\alpha\beta}$ must vanish, we conclude that
the non-physical current that we were concerned about, arising from
$G^{\rm S}_{\alpha\beta}$ in \equ{GS}, is exacly cancelled by another
non-physical one arising from the spreading of the bulk WFs.
Thus, in the end, the physical edge current has pure circulating
character and is related only to antisymmetric Cartesian tensors.

\subsection{Reciprocal-space expressions}
The above two final expressions, \eqs{mlc}{mic}, are given in terms of
bulk WFs. Therefore the total orbital magnetization $\M = \M\sLC + \M\sIC$
of the finite sample converges in the thermodynamic limit to a bulk,
boundary-insensitive, material property. Next, using the WF definition,
\equ{wfs}, we are going to transform $\M\sLC$ and $\M\sIC$ into equivalent
expressions involving BZ integrals of Bloch orbitals. Specifically, we are
going to prove the two identities
\beq
\M\sLC = \frac{1}{2 c (2\pi)^3 }\,{\rm Im}\,
 \sum_n \intk \me{\park u_{n\k}}{\times H_\k\,}{\park u_{n\k}}\;,
\label{LC}
\eeq
\beq
\M\sIC = \frac{1}{2 c (2\pi)^3 }\,{\rm Im}\,
\sum_{nn'} \intk E_{n'n\k} \; \me{\park u_{n\k}}{\times}{\park u_{n'\k}}\;.
\label{IC}
\eeq
These two expressions generalize to the multi-band case
our previous finding for the case of a single occupied band.\cite{Thonhauser05}
There is an important difference, however; while in the single-band
case \eqs{LC}{IC} are {\it separately} gauge-invariant, only their
{\it sum} is gauge-invariant in the multi-band case, as we shall
see in Sec.~\ref{sec:gauge}.

We carry the derivation in reverse, starting from \eqs{LC}{IC}
and showing that they reduce to \eqs{mlc}{mic}.  First, using
\equ{wfs}, we get \bea \ket{\park u_{n\k}} &=& -i \sum_\R
\ei{\k \cdot(\R - \r)} (\r - \R) \, \ket{n\R} \;, \nn H_\k \ket{\park u_{n\k}}
&=& -i \sum_\R \ei{\k \cdot(\R - \r)} H (\r - \R) \, \ket{n\R} \;.\label{basics}
\eea Since the velocity operator is $\v = i[H,\r] = i[H,(\r - \R)]$, and
exploiting $(\r - \R) \times (\r - \R) = 0$, we may express \equ{LC} as \beq
\M\sLC = - \frac{1}{2 c \Omega N_c} \sum_{n\R} \me{n\R}{(\r - \R) \times
\v}{n\R} , \label{circc} \eeq where the number of cell $N_c$ here is formally
infinite, and  appears because the $\ket{u_{n\k}}$ are normalized differently
from the WFs. Since we limit ourselves to the case of an insulator with zero
Chern invariant, the WFs enjoy the usual translational symmetry, and
\equ{circc} is indeed identical to \equ{mlc}.

Next, we address \equ{IC}, whose second factor in the
integral is
\bea\lefteqn{\me{\park u_{n\k}}{\times}{\park u_{n'\k}} =}\nn
&=&\frac{1}{N_c} \sum_{\R \R'} \ei{\k \cdot (\R' - \R)} \me{n\R}{\, (\r - \R)
\times (\r - \R') \,}{n'\R'} \nn &=& \frac{1}{N_c} \sum_{\R \R'} \ei{\k \cdot
(\R' - \R)} \, (\R' - \R) \times \me{n\R}{\,\r \,}{n'\R'} ,  \label{BC}
\eea
where the last line follows because only the cross terms survive from
the product $(\r - \R) \times (\r - \R')$.
We then exploit \beq \me{n'\R'}{H}{n\R} = \frac{\Omega}{(2\pi)^3} \intk
\ei{\k \cdot (\R' - \R)} E_{n'n\k} \label{band} \eeq in
order to rewrite \equ{IC} as
\beq \M\sIC = \frac{\rm Im}{2 c \Omega N_c}
\sum_{\substack{\,\R\R'\\mn}} (\R' - \R) \times \me{m\R}{\r}{n\R'}
\me{n\R'}{H}{m\R} . \label{IC3} \eeq
Since the matrix elements only depend on
the {\it relative} WF coordinate $\R' - \R$, \equ{IC3} is transformed into
\beq \M\sIC = \frac{1}{2 c \Omega}\,{\rm Im} \, \sum_{mn\R} \R \times
\me{m\0}{\r}{n\R} \me{n\R}{H}{m\0} . \label{IC4} \eeq Using \equ{donat},
it is then easy to check that \equ{IC4} is indeed identical to \equ{mic}.

This completes our proof. Our final expression for the macroscopic orbital
magnetization of a crystalline insulator is
\bea \M &=& \frac{1}{2 c (2\pi)^3 }\,{\rm Im}\,  \sum_{nn'}
\intk \bra{\park u_{n'\k}} \nn
&&\qquad\times (\, H_\k\, \delta_{nn'} \, + E_{n'n\k} \,)\,
\ket{\park u_{n\k}}\;. \label{M} \eea
Owing to the occurrence of $H_\k$ and $E_{nn'\k}$ with the {\it same}
sign (in contrast to the magnetization of an individual wavepacket discussed
in Ref.~\onlinecite{Sundaram99}), \equ{M} does not appear at first sight
to be invariant with respect to
translation of the energy zero. However, the zero-Chern-invariant
condition---compare \equ{M} to \equ{chern}---enforces such invariance.
As for the gauge invariance of \equ{M}, this will be demonstrated in the
next subsection.

\subsection{Proof of gauge invariance}
\label{sec:gauge}

Here we prove the gauge invariance in the
multi-band sense of the Chern invariant, \equ{chern}, and of our  main
expression for the macroscopic magnetization, \equ{M}. While these
expressions are BZ integrals, we will actually prove that even their {\it
integrands} are gauge-invariant.  To this end, we will show that both
integrands can be expressed as traces
of gauge-invariant one-body operators acting on the Hilbert
space of lattice-periodical functions.

Our key ingredients are the effective Hamiltonian $H_\k$, the ground-state
projector \beq P_\k = \sum_n \ket{u_{n\k}} \bra{u_{n\k}} , \label{proj}
\eeq and its orthogonal complement $Q_\k = 1 - P_\k$. These three operators
are obviously unaffected by any unitary mixing of the $\ket{u_{n\k}}$
among themselves at a given $\k$, and therefore any expression built
only from these ingredients will be a manifestly multi-band gauge-invariant
quantity.  In particular, we define the three quantities
\beq
f\kab= \mbox{tr} \big\{ (\da P_\k) \, Q_\k \, (\db P_\k) \big\} \;,
\label{fdef}
\eeq
\beq
g\kab= \mbox{tr} \big\{ (\da P_\k) \, Q_\k \, H_\k \, Q_\k \, (\db P_\k) \big\} \;,
\label{gdef}
\eeq
\beq
h\kab= \mbox{tr} \big \{ H_\k \, (\da P_\k) \, Q_\k \, (\db P_\k) \big\} \;,
\label{hdef}
\eeq
where $\da = \partial/\partial k_\alpha$ and
the trace is over electronic states.
We are going to show that the Chern invariant and the magnetization
can be expressed as integrals of $f_\k$ and of $g_\k+h_\k$, respectively.

 From \equ{proj} it follows that
\beq
\da P_\k = \sum_n \big( \, \ket{\da u_{n\k}} \bra{u_{n\k}}
   + \ket{u_{n\k}} \bra{\da u_{n\k}} \, \big)
\label{a1}
\eeq
so that
\beq
(\da P_\k) \, Q_\k \, (\db P_\k) = \sum_{nn'} \ket{u_{n\k}} \me{\da
u_{n\k}}{Q_\k}{\db u_{n'\k}} \bra{u_{n'\k}} .
\label{a2}
\eeq
We now specialize to the ``Hamiltonian gauge'' in which the Bloch
functions $\ket{u_{n\k}}$ are eigenstates of $H_\k$ with eigenvalues
$\epsilon_{n\k}$.  Inserting \equ{a2} into \eqs{fdef}{hdef} and using a
similar approach for \equ{gdef}, the three quantities can be
written as
\bea
f\kab &=& \sum_{n} \ev{\da u_{n\k} | \db u_{n\k}}
\nonumber\\
&-& \sum_{nn'} \ev{\da u_{n\k} | u_{n'\k}} \ev{ u_{n'\k} | \db u_{n\k}} ,
\label{f2}
\eea
\bea
g\kab &=& \sum_{n} \ev{\da u_{n\k} | H_\k | \db u_{n\k}}
\nonumber\\
&-& \sum_{nn'} \epsilon_{n'\k} \ev{\da u_{n\k} | u_{n'\k}}
\ev{ u_{n'\k} | \db u_{n\k}} ,
\label{g2}
\eea
and
\bea
h\kab &=& \sum_{n} \epsilon_{n\k} \ev{\da u_{n\k} | \db u_{n\k}}
\nonumber\\
&-& \sum_{nn'} \epsilon_{n\k} \ev{\da u_{n\k} | u_{n'\k}}
\ev{ u_{n'\k} | \db u_{n\k}} .
\label{h2}
\eea

Regarded as 3$\times$3 Cartesian matrices, Eqs.~(\ref{fdef}-\ref{hdef})
are clearly Hermitian, so that the antisymmetric parts of
Eqs.~(\ref{f2}-\ref{h2}) are all pure imaginary.  Thus, the information
content of the antisymmetric part of $f\kab$ is contained in the
gauge-invariant real vector quantity
\beq
\tilde{f}_{\k,\alpha}= -{\rm Im} \,\varepsilon_{\alpha\beta\gamma} \,
f_{\k,\beta\gamma} ,
\label{ftilde}
\eeq
where $\varepsilon_{\alpha\beta\gamma}$ is the antisymmetric tensor.
We define $\tilde{g}_{\k,\alpha}$ and $\tilde{h}_{\k,\alpha}$ in
the corresponding way in terms of $g_{\k,\beta\gamma}$ and
$h_{\k,\beta\gamma}$ respectively.  Looking at the
second term in \equ{f2} and using $\da\ev{u_{n\k} | u_{n'\k}}=
\da\delta_{nn'}=0$, we find that its antisymmetric part
vanishes, and in fact $\f_\k$ is nothing other than the Berry
curvature.  We thus recover the Chern invariant of \equ{chern}
in the form
\beq
{\bf C} = \frac{1}{2\pi} \intk \f_\k .
\label{chern2}
\eeq

Next, inspecting the second terms of \eqs{g2}{h2}, we find that
neither of these terms vanishes by itself under antisymmetrization.
However, the {\it sum} of these two terms does vanish under
antisymmetrization.  Using the sum only, and comparing with
\equ{M}, we find that the magnetization may be written in the
manifestly gauge-invariant form
\beq
\M = \frac{-1}{2 c (2\pi)^3 }\,\intk (\g_\k+\h_\k) .
\label{MM}
\eeq
(The sign reflects the fact that the electron has negative charge.)
This completes the proof that the integrand in \equ{M} is multi-band
gauge-invariant.

Notice that if we take the first term only in \equ{g2} and
antisymmetrize, we get the integrand in $\M\sLC$ (times a multiplicative
constant); the same holds for \equ{h2} and $\M\sIC$. However, the second
terms in \eqs{g2}{h2} have nonzero antisymmetric parts which are essential
to their gauge-invariance. Therefore, $\M\sLC$ and $\M\sIC$ as defined
above are {\it not} separately gauge-invariant, except in the single-band
case.\cite{Thonhauser05}

On the other hand, it is possible to regroup terms and
write $\M=\widetilde{\M}\sLC+\widetilde{\M}\sIC$, where
\beq
\widetilde{\bf M}\sLC = \frac{-1}{2 c (2\pi)^3 }\,\intk \g_\k
\label{MLC2}
\eeq
and
\beq
\widetilde{\bf M}\sIC = \frac{-1}{2 c (2\pi)^3 }\,\intk \h_\k
\label{MIC2}
\eeq
{\it are} individually gauge-invariant, even in the multi-band case.
This raises the fascinating question as to whether these two contributions
to the orbital magnetization are, in principle, independently measurable.
On the one hand, Berry has emphasized in his milestone paper\cite{Berry84}
that any gauge-invariant property should be potentially observable.
On the other hand, any measurement of orbital magnetization---or,
equivalently, of dissipationless macroscopic surface currents---will
only be sensitive to their sum.  At the present time we have no
insight into how to propose an experiment that could
distinguish them, and we therefore leave this as an open question.

In Appendix A, we show how to compute $\f_\k$, $\g_\k$, and $\h_\k$
on a 3D k-mesh using finite-difference methods to approximate the
derivatives in Eqs.~(\ref{fdef}-\ref{hdef}).

\subsection{Heuristic extension to metals and Chern insulators}
\label{sec:metalchern}

All of the above results are derived under the hypothesis
that the crystalline system is a KS insulator in which the Chern invariant,
\equ{chern}, is zero. These conditions, in fact, are essential for expressing
any ground-state property in terms of WFs. Nonetheless the {\it integrand} in
our final reciprocal-space expression, \equ{M}, is gauge-invariant. This
suggests a possible generalization to Chern insulators (defined as
insulators with nonzero Chern invariant) and even to KS metals.

We notice that \equ{M} is somehow reminiscent of the Berry-phase formula
appearing in the modern theory of electrical polarization.\cite{KSV,rap-a12}
There is an important difference, however. In the electrical case, the
integrand is {\it not} gauge-invariant, and the formula corresponding to our
\equ{M} only makes sense when integrated over the whole BZ, i.e., for a KS
insulator. Indeed, macroscopic polarization is a well-defined bulk property
only for insulating materials.\cite{rap107} Instead, orbital magnetization is
a phenomenologically well-defined bulk property for both insulating and
metallic materials. Therefore, it is worthwhile to investigate heuristically
the validity of an extension of \equ{M} to the metallic case, even though we
cannot yet provide any formal proof.  Additionally, we also heuristically
investigate Chern insulators. Metals and Chern insulators share the property
that their magnetization has a nontrivial dependence on the chemical
potential $\mu$.

We already observed that \equ{M} is invariant by translation of the energy
zero, but this owes to the facts that the integration therein is performed
over the whole BZ, and that the Chern invariant is zero. If we abandon either
of these conditions, the formula has to be modified in order to restore the
invariance. To this end, we first need to restrict our formulation to the
``Hamiltonian gauge'', where the energy matrix is diagonal:
$E_{nn'\k}=\epsilon_{n\k}\delta_{nn'}$. The $\ket{u_{n\k}}$ are therefore
eigenstates of $H_\k$, and the only gauge freedom allowed is now the
arbitrary choice of their phase.

In the general case, including metals and Chern insulators, we propose
to generalize \equ{M} to
\bea \M &=& \frac{1}{2
c(2\pi)^3 }\, \,{\rm Im}\,  \sum_{n} \intkm \bra{\park u_{n\k}}\nn
&&\qquad\qquad\times (\, H_\k +
\epsilon_{n\k} -2 \mu \,)\,\ket{\park u_{n\k}}\;, \label{M2} \eea
where $\mu$ is the chemical potential (Fermi energy).
\equ{M2} has the desirable invariance property
addressed above. Furthermore, in the metallic case,
\equ{M2} provides a magnetization dependent on $\mu$, as it should.
In the insulating case, when $\mu$ is
varied in the gap, $\M$ changes linearly only if the Chern invariant is nonzero,
and remains constant otherwise.  In fact, \eqs{chern}{M2} imply that
\beq \frac{d \M}{d \mu} = - \frac{1}{c(2\pi)^2 } {\bf C}
\label{dM} \eeq
for any insulator and $\mu$ in the gap.

The modification from \equ{M} to \equ{M2} is the minimal one
enjoying the desired properties. Furthermore, in the single-band case
it is essentially identical to a
formula recently proposed by Niu and coworkers,\cite{Xiao05} whose derivation
rests upon semiclassical wavepacket dynamics. 
We provide strong numerical evidence that this formula retains its
validity well beyond the semiclassical regime, and is in fact the exact
quantum-mechanical expression for the orbital magnetization (in a vanishing
macroscopic ${\bf B}$ field).

An expression related---though not identical---to \equ{M2} occurs in the
theory of the Hall effect. Upon replacement of the quantity in parenthesis
with the identity, one obtains something proportional to the integral
of the Berry curvature over occupied portions of the BZ.
This quantity corresponds to the entire Hall conductivity in quantum-Hall
systems~\cite{Thouless82,Kohmoto85} (which are in fact two-dimensional Chern
insulators\cite{rap127}) and the so-called ``anomalous'' Hall term in metals
with broken TR symmetry. The theory of the anomalous Hall effect
has attracted much attention in the recent
literature.\cite{Jungwirth02,Yao04,Haldane04}

\subsection{The two-dimensional case}
\label{sec:twod}
In two dimensions, the magnetization is a pseudoscalar $M$, and the Chern
invariant is the Chern number $C$ (a dimensionless integer)\cite{Thouless}. 
Our heuristic formula, \equ{M2}, then becomes \bea M &=& \frac{1}{2
c(2\pi)^2}\, \, {\rm Im}\,  \sum_{n} \intkm \bra{\park u_{n\k}}\nn
&&\qquad\qquad\times (\, H_\k \,  + \epsilon_{n\k} -2 \mu \, )\,\ket{\park
u_{n\k}}\; .  \label{M2d} \eea The two-dimensional analogue of \equ{dM} is
\beq \frac{d M}{d \mu} = - \frac{C}{2 \pi c} , \label{dM2d} \eeq The physical
interpretation of this equation is best understood in terms of the chiral
edge states of a finite sample cut from a Chern insulator.  Owing to the main
equation ${\bf \nabla} \times \M = \j /c$, a macroscopic current of intensity
$I = cM$ circulates at the edge of any two-dimensional uniformly magnetized
sample, hence \equ{dM2d} yields \beq \frac{d I}{d \mu} = - \frac{C}{2\pi}
\label{dI} . \eeq This is just what is to be expected: raising the chemical
potential by $d\mu$ fills $dk/2\pi$ states per unit length, i.e.,
$dI=-v\,dk/2\pi$; but the group velocity is just $v=d\mu/dk$.  Thus, \equ{dI}
follows with the interpretation that $C$ is the excess number of chiral edge
channels of positive circulation over those with negative circulation. 
Remarkably, the above equations state that the contribution
of edge states is indeed a bulk quantity, and can be evaluted in the
thermodynamic limit by adopting periodic boundary conditions where the
system has no edges. As already observed, this feature may look
counterintutitive, but a similar behavior has been known for more than 20
years in the theory of the quantum-Hall
effect.\cite{Halperin82,Thouless82,Kohmoto85,Yosh} 

In contrast to our case, a magnetic field is usually present in the
standard theory of the quantum-Hall effect, although it is not strictly
needed.\cite{Haldane88} The role of chiral edge states is
elucidated, for example,\cite{Halperin82,Yosh} by considering a vertical
strip of width $\ell$, where the currents at the right and left boundaries
are $\pm I$. The net current vanishes insofar as $\mu$ is constant throughout
the sample.  When an electric field $\mathcal{E}$ is applied across the
sample, the right and left chemical potentials differ by $\Delta \mu =
\mathcal{E} \ell$ and the two edge currents no longer cancel.  Our \equ{dI}
is consistent with the known quantum-Hall results. In fact, according to
\equ{dI}, the net current is $\Delta I \simeq - C \Delta \mu / 2\pi$, while
the transverse conductivity is defined by $\Delta I = \sigma_{\rm T}
\mathcal{E} \ell$.  We thus arrive at $\sigma_{\rm T} = - C/2\pi$ (or, in
ordinary units, $\sigma_{\rm T} = - Ce^2/h$), which is indeed a celebrated
result.\cite{Thouless,Thouless82,Kohmoto85,Niu85} We stress that the Chern
number $C$ is a bulk property of the system, and can be evaluated by adopting
toroidal boundary conditions, where the edges appear to play no role.

\section{Numerical tests}
\label{sec:nummerics}

In a previous paper\cite{Thonhauser05} we tested \equ{M2d} for the insulating
$C = 0$ single-band case on the Haldane model Hamiltonian,\cite{Haldane88}
described below (Sec.~\ref{sec:Haldane}). In this
special case, \equ{M2d} is {\it not} heuristic, since we provided an
analytical proof. We addressed finite-size realizations of the Haldane model,
cut from the bulk; our analysis confirmed that $M\sLC$ arises entirely from the
magnetization of bulk WFs in the thermodynamic limit, whereas
$M\sIC$ arises from current-carrying surface WFs. Both terms
have also been evaluated in terms of bulk Bloch orbitals, by means of
\equ{M2d}, confirming that the orbital magnetization is indeed a genuine
bulk quantity.

Here we extend this program of checking the correctness of our analytic
formulas by carrying out numerical tests on our new multi-band formula,
\equ{M}, derived for the $C = 0$ insulating case.  This is done using a
four-band model Hamiltonian on a square lattice as described below
(Sec.~\ref{sec:square}).  Furthermore, we perform computer experiments
to assess whether our hypothetical \equ{M2d}, proposed to cover
also the metallic and the $C \neq 0$ insulating cases, is consistent
with calculations on finite samples.  We do this for metals in
Sec.~\ref{sec:metal} using the same square lattice as in
Sec.~\ref{sec:square}, but at fractional band filling.  We then do this
in Sec.~\ref{sec:Haldane}
for Chern insulators using the Haldane model\cite{Haldane88} in a range
of parameters for which $C \neq 0$.

Numerical implementation of Eqs. (\ref{M}), (\ref{M2}), and (\ref{M2d})
is quite
straightforward once one has in hand an efficient method for
evaluating the $\k$-derivatives of the Bloch orbitals.
There are several possible approaches to doing this.
One possibility is to evaluate $\ket{\da u_{n\k}}$ by summation
over states as
\beq
   \ket{\da u_{n\k}} = \sum_{m \neq n} \ket{u_{m\k}}
   \frac{\me{u_{m\k}}{\v_\alpha}{u_{n\k}}}
   {\epsilon_{m\k} - \epsilon_{n\k}}. \label{sumoverstates}
\eeq
This is very practical in the context of tight-binding calculations,
where the sum over conduction bands runs only over a small number of
terms, and we adopted this for the test-case calculations
reported below.
However, in first-principles calculations the sums over conduction
states can be quite tedious, and one has to be careful to use the
correct form for the velocity operator in the matrix elements
(see discussion following \equ{group}).
Alternatively, the needed derivatives of $\ket{u_{n\k}}$ can be obtained
from finite difference methods by making use of the discretized covariant
derivative\cite{Sai02,Souza04} as discussed in Appendix A.
It may also be possible to use standard linear-response
methods\cite{Baroni01}
to compute $\ket{\da u_{n\k}}$, as this is an operation which is
already implemented as part of computing the electric-field
response in several modern code packages.

\subsection{Normal insulating case} \label{sec:square}
\begin{figure}[t]\begin{center}
  \includegraphics[width=0.6\columnwidth]{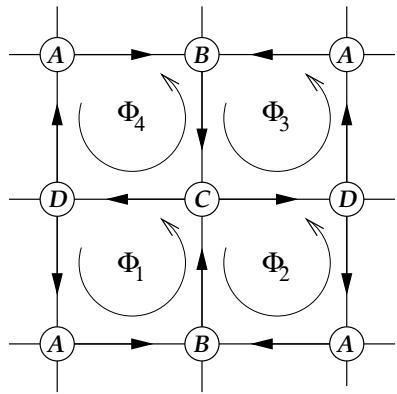}
  \caption{2$\times$2 four-site square lattice used in the numerical tests.
  The absolute value of the hopping parameter $t$ is set to 1.
  $\Phi_{1\cdots4}$ are the threading fluxes through the four plaquettes.}
  \label{fig:square_lattice}
\end{center}\end{figure}

We present in this section numerical tests using a nearest-neighbor
tight-binding Hamiltonian on a 2$\times$2 square lattice in which the
primitive cell comprises four plaquettes, as shown in
Fig.~\ref{fig:square_lattice}. This results in a four-band model.  The modulus
$t$ of the (complex) nearest-neighbor hopping amplitude
is set to 1, thus fixing the energy scale.
TR breaking is achieved by endowing some of the hopping amplitudes with a
complex phase factor $e^{i\phi}$.  This amounts to threading a pattern
of magnetic fluxes through the interiors of the four plaquettes,
as shown in Fig.~\ref{fig:square_lattice}, in such a way that the
threading flux $\Phi_i$ is just the sum of the phase factors associated
with the four bonds delineating plaquette $i$, counted with positive
signs for counterclockwise-pointing bonds and minus signs for
clockwise ones.  The constraint of vanishing macroscopic magnetic field
corresponds to $\Phi_1 + \Phi_2 + \Phi_3 + \Phi_4 = 2\pi\;\times$ integer.
We found that not all flux patterns break TR symmetry.  For instance,
for the flux patterns $\left(\Phi_1,\Phi_2,\Phi_3,\Phi_4\right)=
\left(+\phi,+\phi,-\phi,-\phi\right)$ and
$\left(+\phi,-\phi,+\phi,-\phi\right)$, TR symmetry is restored by some
spatial symmetry (an additional translational symmetry and a mirror symmetry,
respectively); the orbital magnetization then vanishes for any value of
the parameter $\phi$.  On the other hand, the flux pattern
$\left(2\phi,-\phi,0,-\phi\right)$ violates inversion and mirror symmetry,
and therefore realizes TR symmetry breaking.

\begin{figure}[b]\begin{center}
  \includegraphics[width=0.6\columnwidth,angle=270]{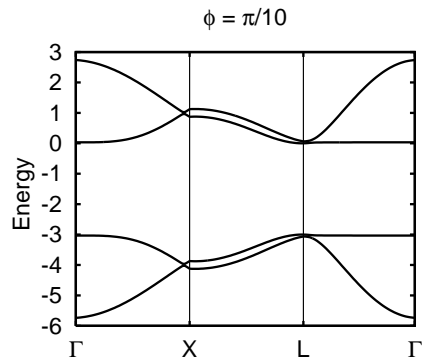}
  \caption{Band structure of the square lattice for $\phi=\pi/10$.
  The flux pattern is
  $\left(\Phi_1,\Phi_2,\Phi_3,\Phi_4\right)=\left(2\phi,-\phi,0,-\phi\right)$,
  and the on-site energies are $\left(E_A,E_B,E_C,E_D\right)=(-3,0,-3,0)$
  (see also Fig.~\ref{fig:square_lattice}).
  The two lower bands are treated as occupied.}
  \label{fig:square_bands}
\end{center}\end{figure}

The on-site energies $\left(E_A,E_B,E_C,E_D\right)$ have been set to the
values $(-3,0,-3,0)$.  This choice results in an insulator with two groups of
two entangled bands as shown in
Fig.~\ref{fig:square_bands}. Switching on the fluxes splits
the bands along the X--L line, which are otherwise two-fold degenerate.  The
$\k$-derivative of Bloch orbitals was computed by the sum-over-states formula
\equ{sumoverstates}.  We treated the two lowest bands as filled and we
verified that the multi-band Chern number is zero.

We built square finite samples, cut from the bulk, made of $L \times L$
four-site unit cells and having $2 L + 1$ sites on each edge. Their orbital
magnetization (dipole per unit area) $M(L)$ is straightforwardly computed
as in \equ{total}. We expect the $L \rightarrow \infty$ asymptotic behavior
\beq M(L) = M + a/L + b/L^2, \eeq where $M$ is the bulk magnetization
according to \equ{M2d}. The terms $a/L$ and
$b/L^2$ account for edge and corner corrections, respectively.

\begin{figure}[t]\begin{center}
  \includegraphics[width=0.6\columnwidth,angle=270]{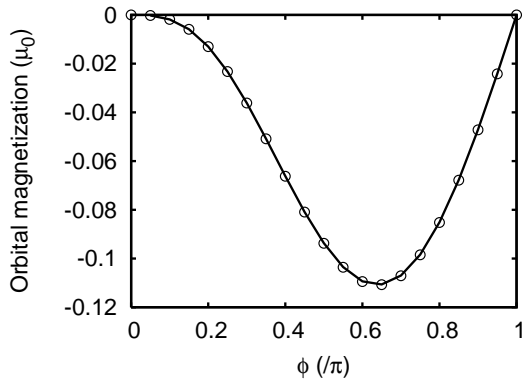}
  \caption{Orbital magnetization of the square-lattice model as a function
  of the parameter $\phi$. The two lower bands are treated as occupied.
  Open circles: extrapolation from finite-size samples. Solid line:
  discretized $\k$-space formula, \equ{M}.}
  \label{fig:square_magn}
\end{center}\end{figure}

We performed calculations up to $L=14$ (841 lattice sites).  The resulting
orbital magnetization as a function of the parameter $\phi$ is shown in
Fig.~\ref{fig:square_magn}.  We independently computed the bulk orbital
magnetization $M$ from a discretization of the reciprocal-space formula
\equ{M2d}. We get well converged results (to within 0.1\%) for a
50$\times$50 $\k$-point mesh in the full BZ.

So far, we have studied a model multi-band insulator, having zero Chern
number. For this specific case we provided above a solid analytic proof of
our reciprocal-space formula, which holds in the thermodynamic limit. Indeed,
the numerical results confirm the correctness of the $\k$-space formula,
while also providing some information
about actual finite-size effects and numerical convergence.

\subsection{Metallic case} \label{sec:metal}

In the previous section we addressed the case of a TR-broken multi-band
insulator, by treating the two lowest bands as occupied. Here we are
going to extend our analysis to the metallic case. We are using the same
model Hamiltonian as in the previous section, but we allow the Fermi level to
span the energy range roughly from $-$5.45 to 2.45 energy units, namely
from the bottom of the lowest band to the top of the highest one.
In order to smooth Fermi-surface singularities, and to obtain well converged
results, we adopt the simple Fermi-Dirac smearing technique, widely used in
first-principle electronic-structure calculations. This amounts to replace,
the (integer) Fermi occupation factor $\Theta(\mu - \epsilon_{n\k})$ with a
suitable smooth function $f_\mu(\epsilon_{n\k})$. We therefore replace in
\equ{M2d}: \beq \sum_{n,\epsilon_{n\k}<\mu} \rightarrow \sum_n
f_\mu(\epsilon_{n\k}). \eeq Reasoning in terms of a fictitious temperature,
one may choose a Fermi-Dirac distribution \beq f_\mu(\epsilon) =
\frac{1}{1+\exp[(\epsilon-\mu)/\sigma]} . \label{FD} \eeq In all subsequent
calculations, we set $\sigma = 0.05$ a.u., which provides good convergence.

We compute the orbital magnetization as a function of the chemical potential
$\mu$ with $\phi$ fixed at $\pi/3$.  Using the same procedure as in the
previous section, we compute the orbital magnetization by the means of the
heuristic $\k$-space formula, \equ{M2d}, and we compare it to the
extrapolated value from finite samples, from $L$=8 (289 sites) to $L$=16 (1089
sites). We verified that a $\k$-point mesh of 100$\times$100 gives well
converged results for the bulk formula, \equ{M2d}.

\begin{figure}\begin{center}
  \includegraphics[width=0.6\columnwidth,angle=270]{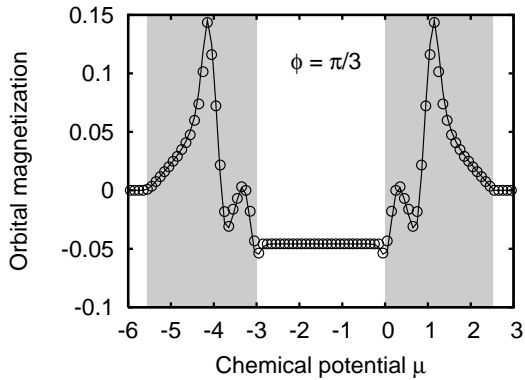}
  \caption{Orbital magnetization of the square-lattice model as a function
  of the chemical potential $\mu$ for $\phi = \pi/3$. The shaded areas
  correspond to the two groups of bands.
  Open circles: extrapolation from finite-size samples. Solid line:
  discretized $\k$-space formula, \equ{M2d}.}
  \label{fig:metal_magn}
\end{center}\end{figure}

The orbital magnetization as a function of the chemical potential for
$\phi=\pi/3$ is shown in Fig.~\ref{fig:metal_magn}.
The resulting values agree to a good level, and provide solid numerical
evidence in favor of \equ{M2d}, whose analytical proof is still lacking.
The orbital magnetization initially increases as the filling of the lowest
band increases, and rises to a maximum at a $\mu$ value of about $-$4.1.
Then, as the filling increases, the first (lowest) band crosses the second
band and the orbital magnetization decreases, meaning that the two bands
carry opposite-circulating currents giving rise to opposite contributions
to the orbital magnetization. The orbital magnetization remains constant
when $\mu$ is scanned through the insulating gap. Upon further increase
of the chemical potential, the orbital
magnetization shows a symmetrical behavior as a function of $\mu$, the two
upper bands having equal but opposite dispersion with respect to the two
lowest bands (see Fig.~\ref{fig:square_bands}).

\subsection{Chern insulating case} \label{sec:Haldane}

In order to check the validity of our heuristic \equ{M2d} for a Chern
insulator, we switch to the Haldane model Hamiltonian\cite{Haldane88} that
we used in a previous paper\cite{Thonhauser05} to address the $C = 0$
insulating case. In fact, depending on the parameter choice, the Chern
number $C$ within the model can be either zero or nonzero (actually, $\pm
1$).

\begin{figure}\begin{center}
  \includegraphics[width=0.8\columnwidth]{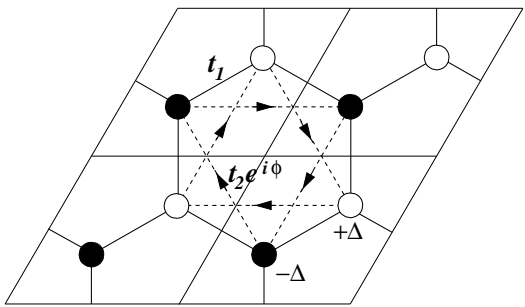}
  \caption{Four unit cells of the Haldane model.
  Filled (open) circles denote sites with $E_0=-\Delta$ ($+\Delta$).
  Solid lines connecting nearest neighbors indicate a real hopping
  amplitude $t_1$; dashed arrows pointing to a second-neighbor site
  indicates a complex hopping amplitude $t_2e^{i\phi}$.
  Arrows indicate sign of the phase $\phi$ for second-neighbor hopping.}
  \label{fig:haldane}
\end{center}\end{figure}

The Haldane model is comprised of a honeycomb lattice with two tight-binding
sites per cell with site energies $\pm \Delta$, real first-neighbor hoppings
$t_1$, and complex second-neighbor hoppings $t_2e^{\pm i\varphi}$, as shown
in Fig.~\ref{fig:haldane}. The resulting Hamiltonian breaks TR
symmetry and was proposed (for $C = \pm 1$) as a realization of the quantum
Hall effect in the absence of a macroscopic magnetic field. Within this two-band
model, one deals with insulators by taking the lowest band as occupied.

In our previous paper~\cite{Thonhauser05} we restricted ourselves to $C = 0$
to demonstrate the validity of \equ{M2d}, which was also analytically proved.
In the present work we address the $C \neq 0$ insulating case, where instead
we have no proof of \equ{M2d} yet.  We are thus performing computer
experiments in order to explore uncharted territory.

Following the notation of Ref.~\onlinecite{Haldane88}, we choose the
parameters $\Delta = 1$, $t_1 = 1$ and $|t_2| = 1/3$. As a function
of the flux parameter $\phi$, this system undergoes a transition from zero
Chern number to $|C|=1$ when $|\sin\phi| > 1/\sqrt{3}$.

\begin{figure}[t]\begin{center}
  \includegraphics[width=0.6\columnwidth,angle=270]{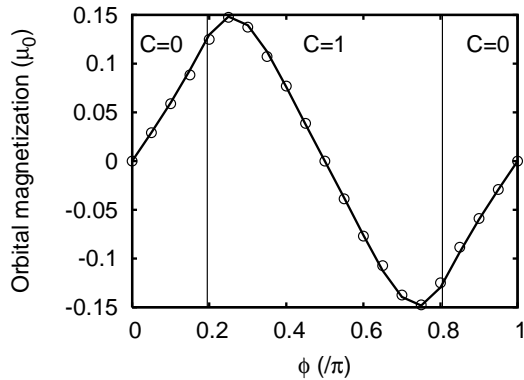}
  \caption{Orbital magnetization of the Haldane model as a function
  of the parameter $\phi$. The lowest band is treated as occupied.
  Open circles: extrapolation from finite size samples.
  Solid line: \equ{M2d}. The system has non-zero Chern number in the region
  in between the two vertical lines.}
  \label{fig:chern_magn}
\end{center}\end{figure}

First we checked the validity of \equ{M2d} in the Chern insulating case by
treating the lowest band as occupied. We computed the orbital magnetization as a
function of $\phi$ by \equ{M2d} at a fixed $\mu$ value, and we compared it to
the magnetization of finite samples cut from the bulk.  For the periodic system,
we fix $\mu$ in the middle of the gap; for consistency, the finite-size
calculations are performed at the same $\mu$ value, using the Fermi-Dirac
distribution of \equ{FD}. The finite systems have therefore fractional orbital
occupancy and a noninteger number of electrons. The biggest sample size
was made up of 20$\times$20 unit cells (800 sites).
The comparison between the finite-size extrapolations and the discretized
$\k$-space formula is displayed in Fig.~\ref{fig:chern_magn}. This heuristically
demonstrates the validity of our main results, \eqs{M2}{M2d}, in the
Chern-insulating case.

\begin{figure}[t]\begin{center}
  \includegraphics[width=0.6\columnwidth,angle=270]{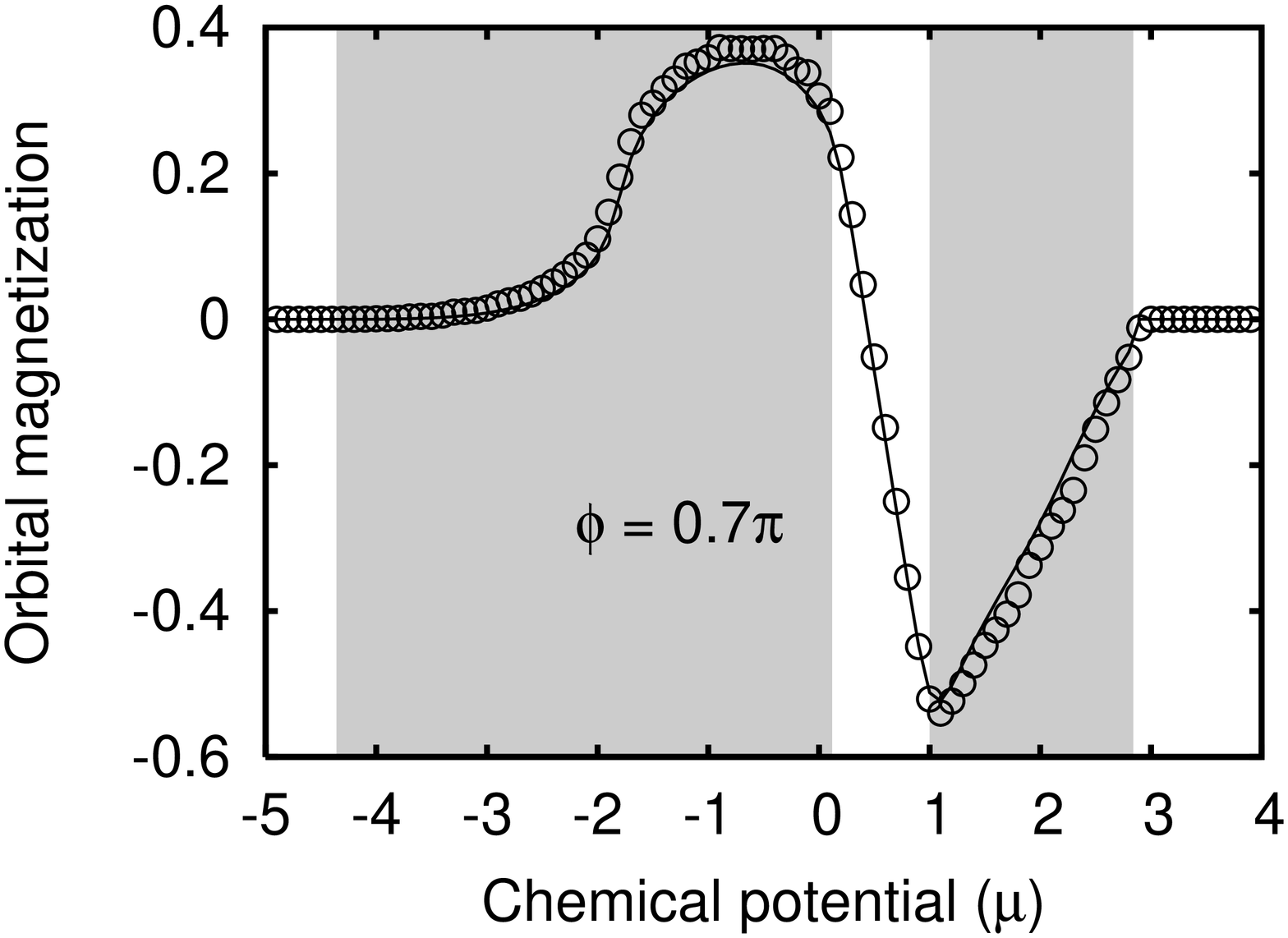}
  \caption{Orbital magnetization of the Haldane model as a function
  of the chemical potential $\mu$ for $\phi = 0.7\pi$. The shaded areas
  correspond the position of the two bands.
  Open circles: extrapolation from finite-size samples. Solid line:
  discretized $\k$-space formula, \equ{M2d}.}
  \label{fig:chern_metal}
\end{center}\end{figure}

Next, we checked the validity of \equ{M2d} for the most general case,
following the transition from the metallic phase to the Chern
insulating phase as a function of the chemical potential $\mu$. To this
aim we keep the model Hamiltonian fixed, choosing  $\phi = 0.7\pi$; for
$\mu$ in the gap this yields a Chern insulator. The behavior of the
magnetization while $\mu$ varies from the lowest-band region, to the
gap region, and then to the highest-band region is displayed in
Fig.~\ref{fig:chern_metal}, as obtained from both the finite-size
extrapolations and the discretized $\k$-space formula. This shows once
more the validity of our heuristic formula. Also notice that in the gap
region the magnetization is perfectly linear in $\mu$, the slope being
determined by the lowest-band Chern number according to \equ{dM2d}.

\section{Conclusions}
\label{sec:conclusions}

We present here a formalism for the calculation of the orbital
magnetization in extended systems with broken TR symmetry,
in the case of vanishing (or commensurate) macroscopic $B$ field.
This extends our previous work of Ref.~\onlinecite{Thonhauser05} to
the multi-band case, essential for realistic calculations.

First, we consider the case of zero Chern invariant, where we provide an an
analytic proof, based upon the Wannier representation. Our main result,
\equ{M}, takes the form of a BZ integral of a gauge-invariant quantity, which
can easily be computed using reciprocal-space discretization. We provide
numerical tests for a two-dimensional model, where our discretized formula
is checked against calculations performed for finite samples cut from the
bulk, with ``open'' boundary conditions. Our numerical tests appear to confirm
that indeed \equ{M} is the correct expression for the orbital magnetization
in a periodic system.

Second, we propose a heuristic extension of \equ{M} to the case of non-zero
Chern invariant, based on the observation that the {\it integrand} in \equ{M}
is gauge invariant, contrary to the analogous electrical case, where only the
BZ integral is gauge-invariant, {\it not} the integrand.\cite{KSV,rap-a12} On
the basis of general considerations (namely, invariance by translation of the
energy zero), the minimal modification extending \equ{M} to the
nonzero-Chern-number case yields \equ{M2}.  Remarkably, \equ{M2} is
essentially identical to a recent expression derived by Xiao \emph{et
al.}~\cite{Xiao05} in the context of a semiclassical approximation. 
We check the full quantum-mechanical validity of \equ{M2} on
a two-dimensional model by means of numerical tests, comparing to finite size
calculations as above. The agreement is excellent, thus providing strong
support for our formula, well beyond the semiclassical regime, even though we
cannot yet provide an analytic proof of it.

Third, since our heuristic \equ{M2} is well-defined for a KS metal,
we also check the validity of \equ{M2} using the same
two-dimensional model as for the metallic case, this time allowing
the chemical potential $\mu$ to be varied through the bands. Once
more the agreement is excellent, providing a numerical
demonstration of the validity of \equ{M2}.

The electrical analogue of the present theory is the modern theory of
polarization,\cite{KSV,rap-a12} developed in the 1990s, and valid for
insulators only. When comparing that theory with the present one, in the
insulating case, there is an important difference which is worth stressing. 
In the electrical case, the whole electronic contribution to the macroscopic
polarization {\it can} be expressed in terms of the electric dipoles of the
bulk WFs. This has a precise counterpart here, where the local-circulation
contribution can in fact be expressed in terms of the magnetic dipoles of the
bulk WFs.  However, we have shown that in the magnetic case there is an
additional ``itinerant-circulation'' contribution
which has {\it no} electrical
analogue. When analyzing finite samples, the latter contribution appears to
be due to chiral currents circulating at the sample boundaries.  Nonetheless,
one of our major findings is that even this contribution can be expressed as
a bulk, boundary-insensitive term.

Both our original expression, \equ{M}, and its heuristic extension,
\equ{M2}, for the orbital magnetization of a crystalline solid can easily
be implemented in existing first-principle electronic structure codes, making
available the computation of the orbital magnetization in crystals, at
surfaces and in reduced dimensionality solids such as nanowires.

\begin{acknowledgments}
This work was supported by ONR grant N00014-03-1-0570, NSF grant
DMR-0233925, and  grant PRIN 2004 from the Italian Ministry of University and Research.
\end{acknowledgments}

\section*{Appendix A: Finite difference evaluation of the Chern
invariant and magnetization}

Using the definition of the covariant derivative\cite{Sai02,Souza04}
\beq
\ket{\wda u_{n\k}}=Q_\k\,\ket{\da u_{n\k}} ,
\label{cderiv}
\eeq
Eqs.~(\ref{fdef}-\ref{hdef}) can be rewritten as
\beq
f\kab= \sum_n \ev{\wda u_{n\k} \vert \wdb u_{n\k}} ,
\label{fdefc}
\eeq
\beq
g\kab= \sum_n \me{\wda u_{n\k}}{H_\k}{\wdb u_{n\k}} ,
\label{gdefc}
\eeq
\beq
h\kab= \sum_{nn'} E_{nn'\k}\, \ev{\wda u_{n'\k} \vert \wdb u_{n\k}} .
\label{hdefc}
\eeq
We assume that the occupied wavefunctions $\ket{u_{n\k}}$ have been
computed on a regular mesh of k-points, and we let
${\bf b}_1$, ${\bf b}_2$, and ${\bf b}_3$ be the primitive reciprocal
vectors that define the k-mesh.  Then the covariant derivative in
mesh direction $i$ can be defined as
\beq
\ket{\wdi u_{n\k}} = {\bf b}_{i\alpha} \, \ket{\wda u_{n\k}}
\label{wdidef}
\eeq
(sum over $\alpha$ implied).
Inserting this into Eqs.~(\ref{fdefc}-\ref{hdefc}) and
taking the antisymmetric imaginary part as in \equ{ftilde}, we
obtain
\beq
\f_\k= \frac{1}{v}\, \epsilon_{ijl} \, {\bf b}_i
\sum_n {\rm Im} \ev{\wdj u_{n\k} \vert \wdl u_{n\k}} ,
\label{fnew}
\eeq
\beq
\g_\k= \frac{1}{v}\, \epsilon_{ijl} \, {\bf b}_i
\sum_n {\rm Im} \me{\wdj u_{n\k} }{H_\k}{\wdl u_{n\k}} ,
\label{gnew}
\eeq
\beq
\h_\k= \frac{1}{v}\, \epsilon_{ijl} \, {\bf b}_i
\sum_{nn'} E_{nn'\k} \, {\rm Im} \ev{\wdj u_{n'\k} \vert \wdl u_{n\k}} ,
\label{hnew}
\eeq
where a sum over $ijk$ is implied and $v$ is the volume of the
unit cell of the k-space mesh.  On this mesh, the BZ
integral in \equ{chern2} becomes a summation
\beq
{\bf C} = \frac{1}{2\pi} \sum_\k \epsilon_{ijl} \, {\bf b}_i
\sum_n {\rm Im} \ev{\wdj u_{n\k} \vert \wdl u_{n\k}}
\label{chern3}
\eeq
and similarly for the magnetization in \equ{MM}.

The appropriate finite-difference discretization of the covariant
derivative in mesh direction $i$ is\cite{Sai02,Souza04}
\beq
\ket{\wdi u_{n\k}} = \frac{1}{2}\,\Big(\,\ket{\wu\nkpbi}-\ket{\wu\nkmbi} \Big)
\label{wdi}
\eeq
where $\ket{\wu\nkpq}$ is the ``dual'' state, constructed as a
linear combination of the occupied $\ket{u\nkpq}$ at neighboring
mesh point $\q$, having the property that $\ev{u_{n'\k}\vert
\wu\nkpq}=\delta_{n'n}$.  This ensures that $\ev{u_{n'\k}\vert
\wdi u_{n\k}}=0$ consistent with \equ{cderiv}, and is solved by
the construction\cite{Sai02,Souza04}
\beq
\ket{\wu\nkpq}=\sum_{n'}\,(S^{-1}_{\k,\k+\q})_{n'n}\,\ket{u_{n',\k+\q}}
\eeq
where
\beq
(S_{\k,\k+\q})_{nn'} = \ev{ u_{n\k} \vert u_{n',\k+\q} } .
\label{Sdef}
\eeq

Eqs.~(\ref{fnew}-\ref{Sdef}) provide the formulas needed
to calculate the three gauge-invariant quantities $\f_\k$, $\g_\k$,
and $\h_\k$ on each point of the k-mesh.  By summing these as
in \equ{chern3} it is straightforward to obtain $\bf C$,
$\widetilde{\M}\sLC$, and $\widetilde{\M}\sIC$, respectively.
Since we have derived this finite-difference representation using
gauge-invariant quantities at each step, it
is not surprising that we obtain the gauge-invariant contributions
$\widetilde{\M}\sLC$ and $\widetilde{\M}\sIC$, as opposed to the
gauge-dependent $\M\sLC$ and $\M\sIC$, from this approach.
\\[1cm]

\section*{Appendix B: The nonAbelian Berry curvature}
It has been noticed in Sec.~\ref{sec:gauge} that the vector quantity
$\f_\k$ is the Berry curvature. From \eqs{f2}{ftilde}, this can be regarded
as
the trace of the $N_b \times N_b$ matrix $\F_{\k}$ having vector
elements
\bea
\F_{\k,nn'} &=& i \, \ev{\park u_{n\k} | \times | \park u_{n'\k}}
\nonumber\\
&-& i \sum_m \ev{\park u_{n\k} | u_{m\k}} \times \ev{ u_{m\k} |
\park u_{n'\k}} . \label{F}
\eea
This quantity is known within the theory of the geometric phase as the
nonAbelian Berry curvature,~\cite{nonAb} and characterizes the evolution of an
$N_b$-dimensional manifold (here, the states $\ket{u_{n\k}}$) in a parameter
space (here,  $\k$-space). The nonAbelian curvature is gauge-covariant,
meaning
that if the states are unitarily transformed as \beq \ket{u_{n\k}} \;
\rightarrow \; \sum_{n'} U_{nn'}(\k) \ket{u_{n'\k}} , \eeq then the matrix
$\F_{\k}$ transforms as \beq \F_{\k,nn'}  \; \rightarrow \; \sum_{mm'}
U_{nm}^\dagger(\k) \F_{\k,mm'} U_{m'n'}(\k) . \label{covar} \eeq This
implies
that the  invariants of the matrix $\F_{\k}$, such as its trace $\f_\k$, are
gauge-invariant. In fact, as discussed in Sec.~\ref{sec:gauge}, $\f_\k$
behaves like a standard (i.e., Abelian) curvature.

We also notice that the energy matrix $E_\k$, \equ{energy}, is also
gauge-covariant in the sense of \equ{covar}. It is then easy to verify that
the
trace (over the band indices) of the matrix product $E_\k \,\F_\k$ is a
gauge-invariant quantity. In fact, this trace is identical to $\h_\k$ as
defined
in Sec.~\ref{sec:gauge}, whose gauge-invariance we proved in a different
way.
The special $N_b=1$ case was previously dealt with in
Ref.~\onlinecite{Thonhauser05},
where the analogue of $\h_\k$ takes the form of the product of energy times
curvature, both gauge-invariant quantities. The present finding shows that,
in
the multi-band case, this must be generalized as the trace of the (matrix)
product $E_\k$ times $\F_\k$, both gauge-covariant quantities.



\end{document}